\documentclass[aps,prd,preprint,floatfix,superscriptaddress,longbibliography,a4paper]{revtex4-1}
\pdfoutput=1
\usepackage{color}
\usepackage{graphicx}
\usepackage{textcomp}
\usepackage{subfigure}
\usepackage{feynmp}
\usepackage{amsmath,amssymb, array}
\usepackage{enumerate}
\usepackage{slashed}
\usepackage[normalem]{ulem}
\usepackage{hhline}
\usepackage{hyperref}
\usepackage{empheq}
\usepackage{cases}
\hypersetup{colorlinks=true,
    citecolor=blue,
	linkcolor=blue,
	filecolor=magenta,      
	urlcolor=blue}
\usepackage{amsthm} 
\theoremstyle{plain} 
\newtheorem{theorem}{Theorem}[] 
\newtheorem{lemma}[theorem]{Lemma} 
\theoremstyle{definition} 

\newcommand{\mathsym}[1]{{}}

\newcommand{\beqa}{\begin{eqnarray}}
\newcommand{\eeqa}{\end{eqnarray}}
\newcommand{\be}{\begin{equation}}
\newcommand{\ee}{\end{equation}}
\newcommand{\ba}{\begin{array}} 
\newcommand{\ea}{\end{array}}

\begin{document} 
\title{Graph-theoretic determination of massless modes in latticized theory-space models}
\bigskip
\author{Ketan M. Patel}
\email{ketan.hep@gmail.com}
\affiliation{Theoretical Physics Division, Physical Research Laboratory, Navarangpura, Ahmedabad-380009, India}

\begin{abstract}
A graph-theoretic method is introduced for analysing fermion mass spectra in latticized theory-space models, including chain models arising from dimensional deconstruction. Fermion mass terms are mapped to bipartite graphs, with fields as vertices and nonvanishing mass terms as edges. The number of massless modes is shown to be fixed by the cardinality of a maximum matching of the associated graph. Moreover, the wave-function support of these modes is restricted to fields reachable from exposed or unmatched vertices by even-length maximum-matching-alternating paths, as characterised by the Dulmage–Mendelsohn decomposition. These results depend only on the topology of latticized theory space and are independent of model parameters. The method enables a systematic construction of latticized models with prescribed numbers and localisation properties of massless modes.
\end{abstract}

\maketitle
 
\section{Introduction}
\label{sec:intro}
Generating hierarchical couplings or widely separated mass scales within quantum field theory remains a longstanding challenge \cite{Dirac:1937ti,Dirac:1938mt,Hogan:1999wh}. This has motivated numerous proposals, including new symmetries, extra spatial dimensions, and dynamical mechanisms such as dimensional transmutation; see, for example, \cite{Giudice:2013yca,Craig:2022eqo} for reviews. A noteworthy approach along this direction is the introduction of multiple copies of fields in four-dimensional theories, together with a highly specific pattern of interactions inspired by the idea of dimensional deconstruction \cite{Arkani-Hamed:2001nha}. Such constructions consist of a sequence of sites in the so-called theory space, each associated with a global or gauged symmetry group. The sites are populated by appropriately charged fields, while adjacent sites are connected through couplings between these fields or through additional link fields charged under the neighbouring symmetry groups. Consequently, these frameworks are often referred to as ``latticized theory-space models'' or ``chain models''. Several examples of these are outlined later in this article.

Depending on the construction, such models may contain fields that remain massless at least at the leading order in a perturbative perspective. Moreover, the localisation of these modes in theory space is of particular interest for generating hierarchical masses and/or couplings. An important question is whether the existence and localisation of such massless modes arise purely from the geometry of interactions in theory space or instead depend on specific choices of the underlying couplings. We show that whenever these features are determined solely by the structure of the latticized theory space, they can be extracted from an associated graph representation and the relevant graph-theoretic properties. More specifically, we demonstrate that both the number of massless modes and the set of fields with which these modes have non-trivial overlap can be determined entirely from graph theory in a model-independent manner. 

We introduce the necessary graph-theoretic concepts relevant to our analysis and establish a direct correspondence between graph properties and the structural features of latticized theory-space models. For concreteness, we focus on the theories involving fermions; however, the concept may be generalised to include other fields. It is shown that this intriguing connection not only reveals the structural features of chain models, but also provides a systematic framework for constructing lattice structures with a desired number of massless modes.

\section{Preliminaries on bipartite graphs}
\label{sec:graph}
A \emph{bipartite graph} $G(L \cup R,E)$ is defined as a collection of vertices divided into two sets $L$ and $R$, and edges forming a set $E$. Each edge in $E$ connects a vertex in $L$ to a vertex in $R$. For an edge $e=(l_i,r_j) \in E$, we say that $e$ is \emph{incident} to the vertices $l_i \in L$ and $r_j \in R$. The graphs we consider are simple, undirected and do not contain loops \cite{LovaszPlummer1986MatchingTheory}. Moreover, we consider that the number of vertices in $L$ and $R$ are identical, i.e. ${\rm dim}(L) = {\rm dim}(R) \equiv n(G)$, without loss of generality. If such a condition is not satisfied for a graph, an appropriate number of disconnected vertices in either of the sets $L$ or $R$ are to be added.

A \emph{matching} in $G$ is a set $M \subseteq E$ of edges such that no vertex of $L \cup R$ is incident on more than one edge of $M$. The matching of maximum size is referred to as \emph{maximum matching} and the size is denoted by $\nu(G)$. It is obvious that $\nu(G) \leq n(G)$. A matching is called \emph{perfect matching} if $\nu(G)=n(G)$. For $\nu(G) < n(G)$, there are vertices in $L \cup R$ on which no edge of maximum $M$ is incident. Such a vertex is called an \emph{exposed vertex}.

A \emph{vertex cover} of $G$ is a set $C \subseteq L \cup R$ of vertices such that each edge of $G$ is incident on at least one of the vertices in $C$. The number of vertices in the \emph{minimum vertex cover} of $G$ is denoted by $\tau(G)$. The set $C$ can be written as $C = C_L \cup C_R$ with $C_L \subseteq L$ and $C_R \subseteq R$. Clearly, $\tau(G) = {\rm dim}(C_L) + \dim(C_R)$. It is also useful to note that while the choice of maximum $M$ or minimum $C$ is not unique for a given graph, their cardinalities are always unique \cite{LovaszPlummer1986MatchingTheory}.

A useful theorem in graph theory relating the sizes of maximum matching and minimum vertex cover is by K{\"o}nig \cite{GraphsMatrices1931}. 
\begin{theorem} \label{th:Konig}
For any bipartite graph, $\nu(G)=\tau(G)$.
\end{theorem}
Therefore, the task of finding $\tau(G)$ reduces to the task of finding a maximum matching in the case of bipartite graphs. For the latter, several algorithms have been proposed in computational theory. The simplest being the so-called Hungarian method based on the works by K{\"o}nig \cite{GraphsMatrices1931,Konig1936Theorie}, and Egerv\'ary \cite{egervary1931matrixok}, which we outline below.

\begin{figure}[!t]
    \centering
    \subfigure[]{
        \makebox[0.30\linewidth][c]{%
            \includegraphics[
                height=0.28\textheight,
                width=0.29\linewidth,
                keepaspectratio
            ]{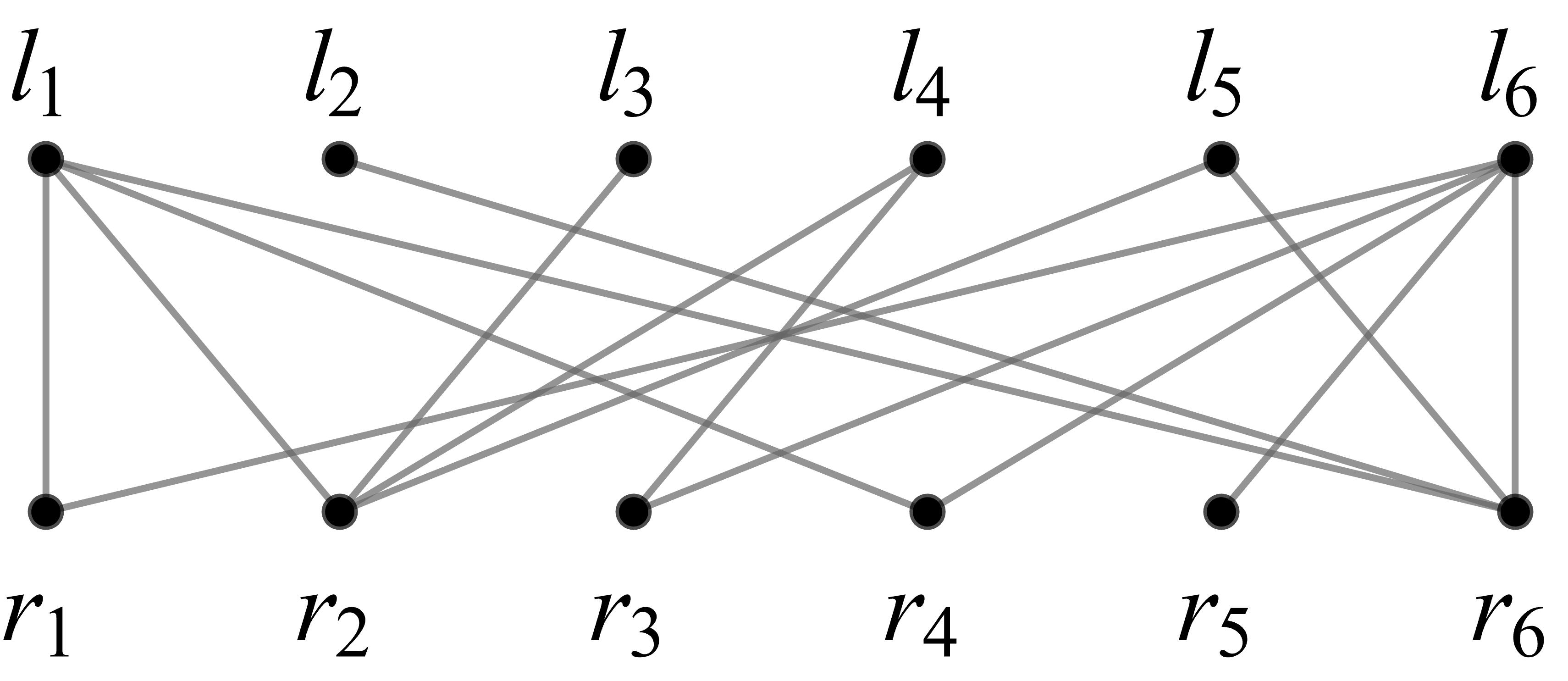}
            \label{fig:demoa}
        }
    }
    \hfill
    \subfigure[]{
        \makebox[0.30\linewidth][c]{%
            \includegraphics[
                height=0.28\textheight,
                width=0.29\linewidth,
                keepaspectratio
            ]{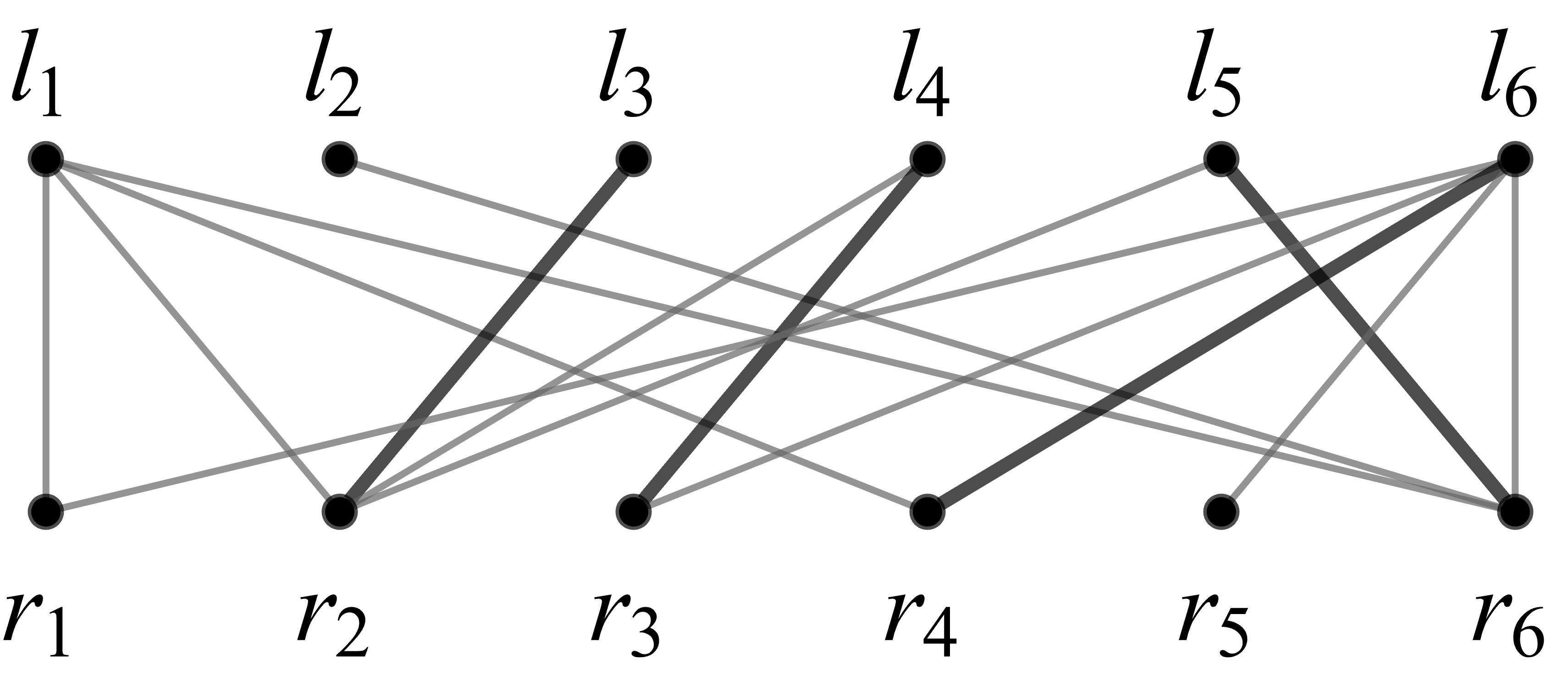}
            \label{fig:demob}
        }
    }
        \hfill
    \subfigure[]{
        \makebox[0.30\linewidth][c]{%
            \includegraphics[
                height=0.28\textheight,
                width=0.29\linewidth,
                keepaspectratio
            ]{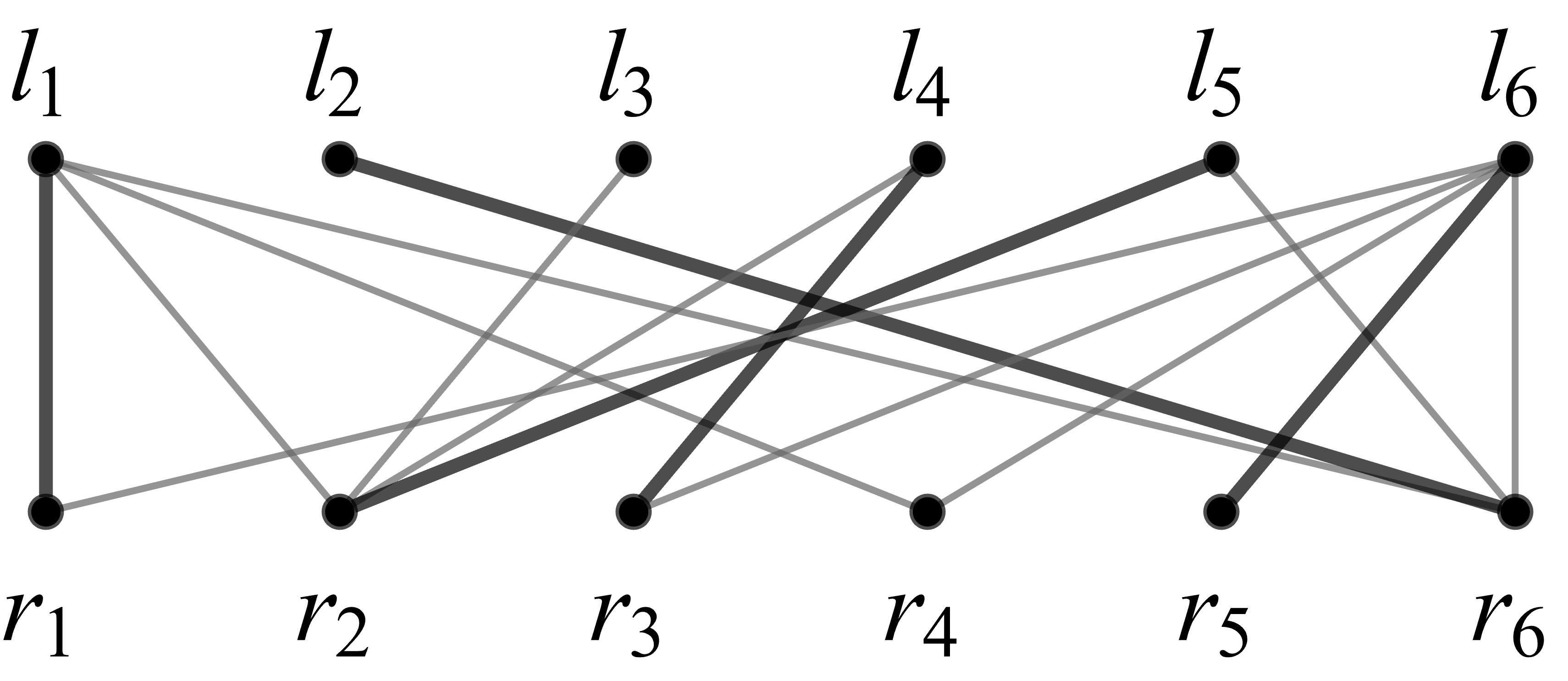}
            \label{fig:demobc}
        }
    }
\caption{Illustration of finding a maximum matching using the Hungarian method for a graph $G$ given in (a). The edges in the matching $M$ are shown by thick lines for the $G$ in (b) and (c). In (b), $M$ is not maximum as an $M$-alternating path, $(l_1,r_4),(r_4,l_6),(l_6,r_1)$, connects the exposed vertices $l_1$ and $r_1$. Updating $M$ gives the new matching shown in (c). The remaining exposed vertices, $l_3$ and $r_4$, have no $M$-alternating path between them, so the $M$ given in (c) is maximum and $\nu(G)=6$.}
\label{fig:demo}
\end{figure}

A \emph{path} $P$ is a collection of edges, e.g. $(l_1,r_1),(r_1,l_3),(l_3,r_6),...,(r_k,l_l),(l_l,r_m)$, where all the vertices are distinct. The edges in $G$ can be divided into two subsets: $M$ and $ E \setminus M = E-M$. An \emph{alternating path with respect to $M$} or an \emph{$M$-alternating path} is a path that alternates between the edges in $M$ and $E \setminus M$. In other words, if an edge $(l_k,r_l) \in M$ then $(r_l,l_m) \in E \setminus M$, if they belong to such a path. An \emph{augmenting path with respect to $M$} is an $M$-alternating path joining two exposed vertices. The following theorem by Berge \cite{Berge1957TwoTheorems} then helps in finding the maximum matching.
\begin{theorem} \label{th:Berge}
$M$ is a maximum matching of $G$ iff there exists no augmenting path with respect to $M$ in $G$.
\end{theorem}
Consequently, one can begin with any $M$ for a given $G$, find an augmenting path $P$ with respect to $M$, and replace $M$ with a new matching which is $(M \setminus P) \cup (P \setminus M)$. This can be continued until no more augmenting path is left. In this way, $\nu(G)$ can directly be computed from graphs \footnote{Software like \textsc{Mathematica} already have such algorithms inbuilt under the function \texttt{FindIndependentEdgeSet}.}. An illustration of this is demonstrated in Fig. \ref{fig:demo}.

\section{Counting massless modes from graph}
\label{sec:count}
In quantum field theory, the mass terms are bilinear in the fields and can be denoted by a graph. For simplicity, consider the case of fermions with Dirac mass terms:
\be \label{Lm}
-{\cal L}_m = \sum_{i,j}\, \mu_{ij}\, \overline{f_L}_i\, f_{Rj}\,+\,{\rm h.c.}\,,\ee
where $f_{L}$, $f_{R}$ are left- and right-chiral components of a Dirac field, respectively. Representing $f_{Li}$ and $f_{Rj}$ as vertices $l_i \in L$ and $r_j \in R$, respectively, a bipartite graph can be constructed which has an edge $(l_i,r_j) \in E$ if $\mu_{ij} \neq 0$. The edge is absent otherwise. For $ n$ copies of fields, the $n \times n$ fermion mass matrix is nothing but a weighted adjacency matrix $A$ of the graph $G$. Namely, 
\be \label{A}
A_{ij} = 
\begin{cases}
  \mu_{ij} & \text{if $(l_i,r_j) \in E$, } \\
  0 & \text{otherwise}.
\end{cases}\ee
Here, $\mu_{ij}$ are generic couplings and no further assumption about their magnitudes or pattern is made. If all the non-vanishing $\mu_{ij}$ are replaced by $1$, then $A$ becomes the standard bi-adjacency matrix of a given bipartite graph \cite{LovaszPlummer1986MatchingTheory}.

The matrix $A$ defined in Eq. (\ref{A}), with its nonvanishing entries treated as generic parameters, is an example of what is known as an Edmonds matrix \cite{Brualdi_Ryser_1991,Murota2010} of the bipartite graph $G$. It is a classical result that the rank of such a matrix equals the size of the maximum matching of the underlying bipartite graph. This was originally established by Edmonds in \cite{Edmonds1967SystemsOD} via a linear-algebraic argument based on transversals and minors. For completeness and to keep the derivation self-contained in the language used below, we now derive this rank–matching correspondence directly using an explicit block decomposition and K{\"o}nig's theorem without relying on Edmonds' original arguments.

Let $C_L = \{l_1,...,l_p\}$ and $C_R = \{r_1,...,r_q\}$ and $C = C_L \cup C_R$ is a minimum vertex cover of the underlying graph $G$. Consequently, $L \setminus C_L = \{l_{p+1},...,l_n\}$ and $R \setminus C_R = \{r_{q+1},...,r_n\}$. This can be arranged without loss of generality since it is possible to relabel the vertices and edges without modifying the structure of the graph. By definition, the cardinality of a minimum vertex cover is at most $n$, since either $L$ or $R$ constitutes a vertex cover: every edge of $G$ is incident to a vertex in $L$ or in $R$. Hence,
\be \label{cond1}
\tau(G) = p + q \le n\,.
\ee

Next, no edge is completely contained in $(L \setminus C_L) \cup (R \setminus C_R)$; otherwise, $C$ is not a vertex cover. Therefore, the matrix $A$ in the aforementioned basis can be written as
\be \label{A_decomp}
A=\left( \ba{cc} \left(A_1\right)_{p\times q} ~~&~~ \left(A_2\right)_{p\times (n-q)} \\
\left(A_3\right)_{(n-p) \times q} ~~&~~ \left(0 \right)_{(n-p) \times (n-q)}\ea \right)\,,\ee
where the subscripts indicate the dimensions of the corresponding blocks. $A$ can be decomposed as
 \be \label{A_decomp2}
A=\left( \ba{cc} \left(A_1\right)_{p\times q} ~~&~~ \left(A_2\right)_{p\times (n-q)} \\
\left(0\right)_{(n-p) \times q} ~~&~~ \left(0 \right)_{(n-p) \times (n-q)}\ea \right)+\left( \ba{cc} \left(0\right)_{p\times q} ~~&~~ \left(0\right)_{p\times (n-q)} \\
\left(A_3\right)_{(n-p) \times q} ~~&~~ \left(0 \right)_{(n-p) \times (n-q)}\ea \right)\,,\ee
and using the fact that ${\rm rank}(X+Y) \leq {\rm rank}(X) + {\rm rank}(Y)$, one finds
\be \label{rankA_2}
{\rm rank}(A) \leq p + {\rm min}(n-p,q)\,.\ee
Condition in Eq. (\ref{cond1}) further simplifies the above to
\be \label{rankA_1}
{\rm rank}(A) \leq p + q = \tau(G)\,.\ee
The rank of $A$ is, therefore, bounded by $\tau(G)$.

Note that ${\rm rank}(A) < \tau(G)$ only if there exist further linear dependencies among the rows or columns of $A_{1,2,3}$. Hence for generic values of $\mu_{ij}$, 
\be \label{rankA}
{\rm rank}(A) = \tau(G)\,.\ee
The above result, along with Theorem \ref{th:Konig}, thus reproduces Edmonds' theorem \cite{Edmonds1967SystemsOD} relevant to our bipartite setting. The generic rank of a matrix with a prescribed zero pattern equals the term rank, i.e., the maximum-matching cardinality of the associated bipartite graph. This immediately gives,
\be \label{res_1}
\text{\emph{number of massless modes}} = n-\nu(G)\,.
\ee
Therefore, the number of massless modes that result due to the topology of the mass terms can be directly read from the graph by finding the cardinality of the maximum matching.

\section{Determining profile of massless modes}
\label{sec:profile}
The composition of massless modes, i.e. their wave-function profile, is also a matter of interest from a model-building perspective. The parameter-independent features of these modes can also be read from graphs.  A useful property of bipartite graphs for this purpose is the Dulmage–Mendelsohn (DM) decomposition \cite{DulmageMendelsohn1958Coverings,DulmageMendelsohn1959Structure}.  Under the DM decomposition, the vertices of a $G(L\cup R, E)$ can be divided into three disjoint sets using its maximum matching $M$.  A vertex is called \emph{even (odd) reachable} if it can be reached by an even (odd) length of $M$-alternating path starting from an exposed vertex in the graph, otherwise it is \emph{unreachable}. Consequently, the subsets ${\cal E} = {\cal E}_L \cup {\cal E}_R$, ${\cal O} = {\cal O}_L \cup {\cal O}_R$ and ${\cal U} = {\cal U}_L \cup {\cal U}_R$ of $L \cup R$ are constructed from the even reachable, odd reachable and unreachable vertices, respectively. Obviously, ${\cal E} \cup {\cal O} \cup {\cal U} = L \cup R$.

The following properties of DM decomposition will be used for deriving the result of our interest.
\begin{lemma} \label{lm:DM1}
${\cal E} \cap {\cal O} = {\cal O} \cap {\cal U} = {\cal U} \cap {\cal E} = \emptyset$  independent of the maximum $M$. 
\end{lemma}
\begin{lemma} \label{lm:DM2}
No edge in $G$ is incident to a vertex in ${\cal E}$ and a vertex in ${\cal E} \cup {\cal U}$.  
\end{lemma}
\begin{lemma} \label{lm:DM3}
In any maximum $M$ of $G$,  every vertex in ${\cal O}$ is matched with a vertex in ${\cal E}$ and every vertex in ${\cal U}$ is matched with another vertex in ${\cal U}$. 
\end{lemma}
The proofs of the above Lemmas can be found in \cite{Irving2006}. Moreover, one can also compute $\nu(G)$ from the cardinalities of ${\cal O}$ and ${\cal U}$ as $\nu(G) = {\rm dim}({\cal O}) + {\rm dim}({\cal U})/2$ \cite{Irving2006}.

Using Lemma \ref{lm:DM1}, the vertices of $G$ can be divided into three disjoint sets. Utilizing the relabelling, we arrange them as ${\cal E}_L = \{l_1,...,l_p\}$, ${\cal O}_L = \{l_{p+1},...,l_{p+q}\}$ and ${\cal U}_L = \{l_{p+q+1},...,l_{n}\}$.  Analogously, ${\cal E}_R = \{r_1,...,r_{p^\prime}\}$, ${\cal O}_R = \{r_{p^\prime+1},...,r_{p^\prime+q^\prime}\}$ and ${\cal U}_R = \{r_{p^\prime+q^\prime+1},...,r_{n}\}$.  The weighted adjacency matrix $A$ defined in Eq. (\ref{A}) can be written in this basis as,
\be \label{A_DM}
A=\left( \ba{ccc} \left(0\right)_{p\times p^\prime} ~~&~~ \left(A_{\cal EO}\right)_{p\times q^\prime} ~~&~~ \left(0\right)_{p\times (n-p^\prime - q^\prime)} \\
\left(A_{\cal OE}\right)_{q \times p^\prime} ~~&~~ \left(A_{\cal OO} \right)_{q \times q^\prime} ~~&~~ \left(A_{\cal OU} \right)_{q \times (n-p^\prime-q^\prime)} \\
\left(0\right)_{(n-p-q) \times p^\prime} ~~&~~ \left(A_{\cal UO} \right)_{(n-p-q) \times q^\prime} ~~&~~ \left(A_{\cal UU} \right)_{(n-p-q) \times (n-p^\prime-q^\prime)} \ea \right)\,.\ee
The zeros in the top-left, top-right and bottom-left blocks follow from Lemma \ref{lm:DM2}.

Consider now an eigenvector $\hat{r}$ of $A$ with vanishing eigenvalue, i.e.,
\be \label{ev_0}
A\, \hat{r} = 0\,\ee
Since $A$ represents the fermion mass matrix, $\hat{r}$ is an eigenvector of a massless mode. In the aforementioned basis, it can be expressed as
\be \label{r0_decomp}
\hat{r} = \left(\hat{r}_{\cal E},\, \hat{r}_{\cal O},\, \hat{r}_{\cal U} \right)^T\,,\ee
where $\hat{r}_{{\cal E},{\cal O},{\cal U}}$ are respectively $p^\prime$, $q^\prime$ and $n-p^\prime-q^\prime$ dimensional column vectors. Eq. (\ref{ev_0}) then leads to
\be \label{ro_cond1}
\sum_{j=1}^{q^\prime} \left(A_{\cal EO} \right)_{ij}\, \left(\hat{r}_{\cal O}\right)_{j} = 0\, \quad \forall i \in \{1, \ldots, p\}\,.\ee
Lemma \ref{lm:DM3} implies that every vertex in ${\cal O}_R$ is matched with a vertex in ${\cal E}_L$ in a maximum matching. This is true only if $p \geq q^\prime$ and at least $q^\prime$ of the total $p$ rows of $A_{\cal EO}$ are non-vanishing. Also, the genericness of couplings implies that these non-vanishing rows are linearly independent. Therefore, ${\rm rank}\left(A_{\cal EO} \right) = q^\prime$ and hence Eq. (\ref{ro_cond1}) must correspond to
\be \label{ro_res1}
 \left(\hat{r}_{\cal O}\right)_{j} = 0\, \quad \forall j \in \{1, \ldots, q^\prime\}\,.\ee
In other words, there are $q^\prime$ variables in $\hat{r}_{\cal O}$ constrained by $p \geq q^\prime$ number of linearly independent equations in (\ref{ro_cond1}) resulting into Eq. (\ref{ro_res1}).

Substituting the above result in Eq. (\ref{r0_decomp}) and using again Eq. (\ref{ev_0}), one finds
\be \label{ru_cond1}
\sum_{j=1}^{n-p^\prime-q^\prime} \left(A_{\cal UU} \right)_{ij}\, \left(\hat{r}_{\cal U}\right)_{j} = 0\, \quad \forall i \in \{1, \ldots, n-p-q\}\,.\ee
Again, Lemma \ref{lm:DM3} implies that $p + q = p^\prime + q^\prime$. It also implies that all the rows and columns of $A_{\cal UU}$ are non-vanishing and linearly independent for generic values of couplings.  $A_{\cal UU}$ is therefore invertible and the last condition reduces to
\be \label{ru_res1}
\left(\hat{r}_{\cal U}\right)_{j} = 0\, \quad \forall j \in \{1, \ldots, n-p^\prime-q^\prime\}\,.\ee
Note that Lemma \ref{lm:DM3} also enforces $q \leq p^\prime$. Therefore,
\be \label{re_cond1}
\sum_{j=1}^{p^\prime} \left(A_{\cal OE} \right)_{ij}\, \left(\hat{r}_{\cal E}\right)_{j} = 0\, \quad \forall i \in \{1, \ldots, q\}\,,\ee
which is obtained after substituting Eqs. (\ref{ro_res1},\ref{ru_res1}) in Eq. (\ref{ev_0}), cannot lead to completely vanishing $\hat{r}_{\cal E}$ in general.

The above implies that $\hat r$ lies entirely within ${\cal E}_R$. An analogous argument, starting from $A^\dagger \hat l = 0$, shows that $\hat l$ is likewise fully contained in ${\cal E}_L$. Consequently, \emph{the massless modes are linear combinations exclusively of fields associated with vertices reachable from exposed vertices by even-length $M$-alternating paths in any maximum matching $M$}. This constitutes another important result of this work, as it characterises the set of fields with non-vanishing overlap with the massless modes purely from the graph \footnote{A \textsc{Mathematica} subroutine for computing the DM decomposition of a bipartite graph is provided
in the ancillary file \texttt{DM\_decomposition.nb} included with
the arXiv submission.}.

It is noteworthy that $A$ given in Eq. (\ref{A_DM}) can be made manifestly block-triangular, which offers a complementary way of viewing the above result. Reorganising the vertices of $L$ as $({\cal E}_L,{\cal U}_L,{\cal O}_L)$ and, those of $R$ as $({\cal O}_R,{\cal U}_R,{\cal E}_R)$, the matrix $A$ can be brought to the lower block-triangular form,
\be \label{A_DM_triang}
A=\left(\ba{ccc}
A_{\cal EO} & 0 & 0\\
A_{\cal UO} & A_{\cal UU} & 0\\
A_{\cal OO} & A_{\cal OU} & A_{\cal OE} \ea\right)\,,\ee
which is the standard DM normal form of the graph \cite{Brualdi_Ryser_1991,Murota2010}. The diagonal blocks are the overdetermined block $A_{\cal EO}$, the square invertible block $A_{\cal UU}$, and the underdetermined block $A_{\cal OE}$. Since $A_{\cal UU}$ is invertible, solving $A\hat r=0$ (or $A^\dagger\hat l=0$) by back-substitution forces the ${\cal U}$- and ${\cal O}$-components to vanish, leaving the null space supported entirely on ${\cal E}$. This reproduces the result derived above by a direct triangular argument.

\section{Extension to Majorana mass terms}
\label{sec:majorana}
The correspondence established above can be extended to include Majorana as well as mixed Dirac-Majorana mass terms with a little modification in the field representation in terms of vertices. The most general mass term of this type for $n_L$ copies of left-chiral and $n_R$ copies of right-chiral fermions is given by
\be \label{mixed_Lm}
-{\cal L}_m = \sum_{a,b=1}^{n_L}\, \sum_{a^\prime,b^\prime=1}^{n_R} \left(\frac{1}{2} (\mu_L)_{ab}\, \overline{f_L}_a\, f_{L b}^c +\frac{1}{2} (\mu_R)_{a^\prime b^\prime}\, \overline{f_R^c}_{a^\prime}\, f_{R b^\prime}+ (\mu_D)_{aa^\prime}\, \overline{f_L}_a\, f_{R a^\prime}\right) + {\rm h.c.}\,, \ee
where the first two terms represent the Majorana mass terms with the usual definition of the charge-conjugated field, $f^c = C \bar{f}^T$, and $C$ is the charge conjugation matrix. The matrices $\mu_L$ and $\mu_R$ are therefore symmetric in flavour space. The set of $n_L + n_R \equiv n$ left- and right-chiral fields is then constructed as $\psi_L = \{f_{L a},f^c_{R a^\prime}\}$ and $\psi_R = \{f^c_{L a},f_{R a^\prime}\}$, respectively. 

As before, one can represent $\psi_{Li}$ and $\psi_{Ri}$ as vertices $l_i \in L$ and $r_i \in R$, respectively, and the edges between them by non-vanishing elements of $\mu_{ij}$, where
\be \label{mixed_mu}
\mu = \left(\ba{cc} \mu_L & \mu_D \\ \mu_D^T & \mu_R \ea\right)\,.\ee 
The graph obtained in this way remains bipartite. Additionally, the symmetric nature of $\mu$ implies $A_{ij} = A_{ji}$ for the corresponding adjacency matrix. Therefore, the underlying graph is symmetric with respect to the interchange $l_i \leftrightarrow r_i$ in this case. From the structure of $\psi_{L,R}$, it is obvious that any permutation or relabelling of the vertices must be identical in the sets $L$ and $R$. Any such operation maintains the symmetric nature of the graph. 

Both main results derived earlier extend verbatim to graphs corresponding to mixed Dirac–Majorana mass terms. The proof in section \ref{sec:count} applies with the additional constraints $A_1^T=A_1$ and $A_3=A_2^T$. The steps in Eqs. (\ref{A_decomp}-\ref{res_1}) remain unchanged, so the number of massless modes is given by $n-\nu(G)$. Likewise, the results of section \ref{sec:profile} extend with the restrictions $A_{\cal O E}=A^T_{\cal E O}$ and $A_{\cal O U}=A^T_{\cal U O}$, while $A_{\cal OO}$ and $A_{\cal UU}$ are symmetric. These conditions do not affect the generality of the proof. Hence, the massless-state eigenvector $\hat{r}$ lies entirely within ${\cal E}_R$. Moreover, by construction, ${\cal E}_L={\cal E}_R$, ${\cal O}_L={\cal O}_R$, and ${\cal U}_L={\cal U}_R$, consistent with the fact that the symmetric nature of $A$ implies $\hat{l}=\hat{r}^*$.

\section{Examples and application}
\label{sec:ex}
We consider a few well-known examples of models with specific arrangements of mass terms among the fields, and analyse their structural features using graphs. The focus is on two main properties derived above: the number and profile of massless modes. 

\begin{figure}[!t]
    \centering
    \subfigure[~Minimal seesaw \cite{King:1999mb}]{
        \makebox[0.48\linewidth][c]{%
            \includegraphics[
                height=0.28\textheight,
                width=0.24\linewidth,
                keepaspectratio
            ]{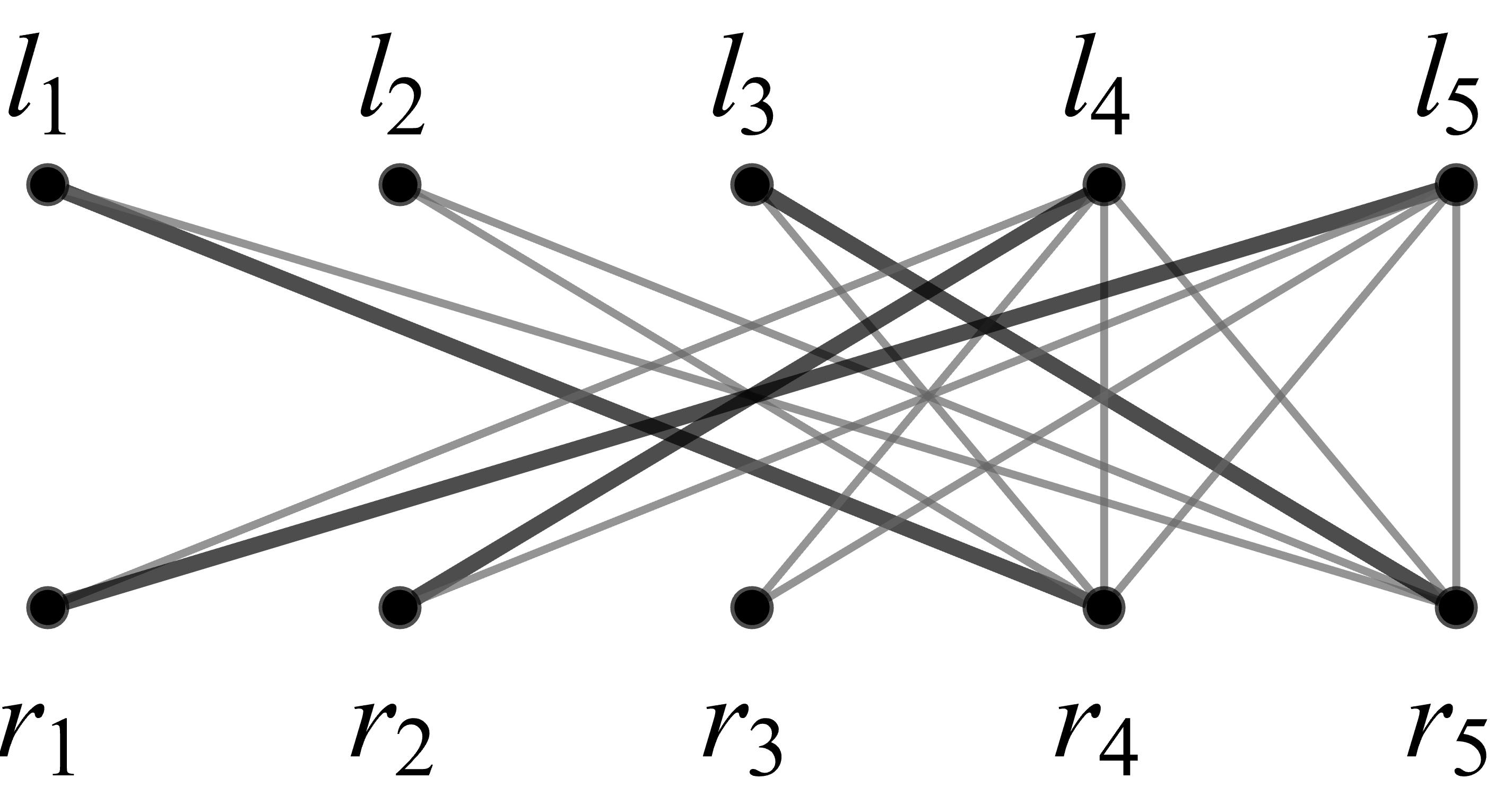}
            \label{fig:seesaw}
        }
    }
    \hfill
    \subfigure[~Clockwork \cite{Giudice:2016yja}]{
        \makebox[0.48\linewidth][c]{%
            \includegraphics[
                height=0.28\textheight,
                width=0.42\linewidth,
                keepaspectratio
            ]{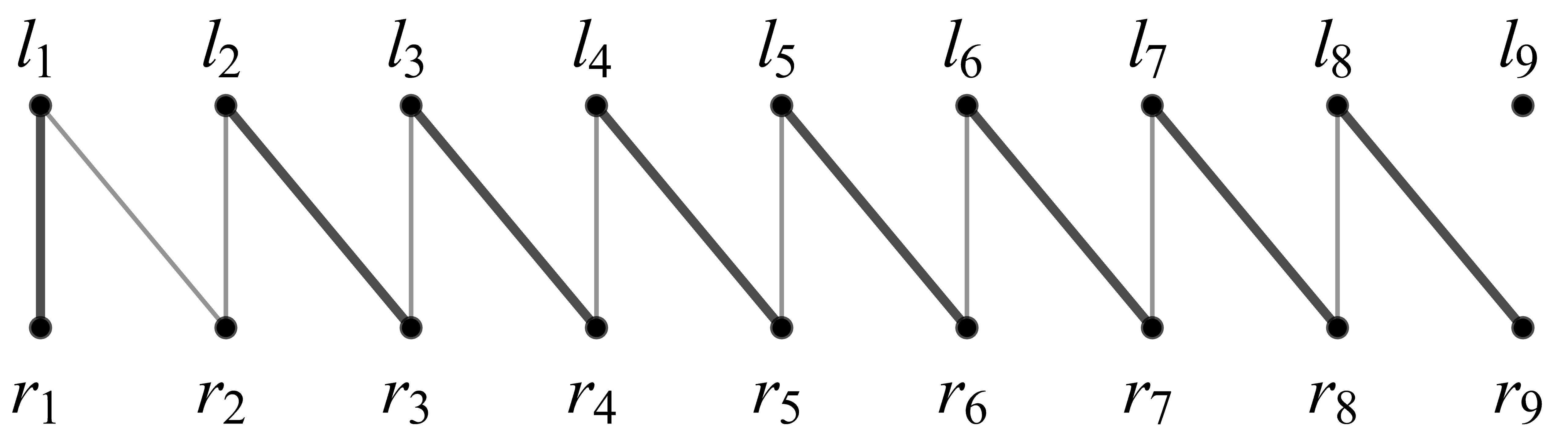}
            \label{fig:clockwork}
        }
    }
    \vspace{0.3cm}
    \subfigure[~Anderson \cite{Craig:2017ppp}]{
        \makebox[0.48\linewidth][c]{%
            \includegraphics[
                height=0.28\textheight,
                width=0.38\linewidth,
                keepaspectratio
            ]{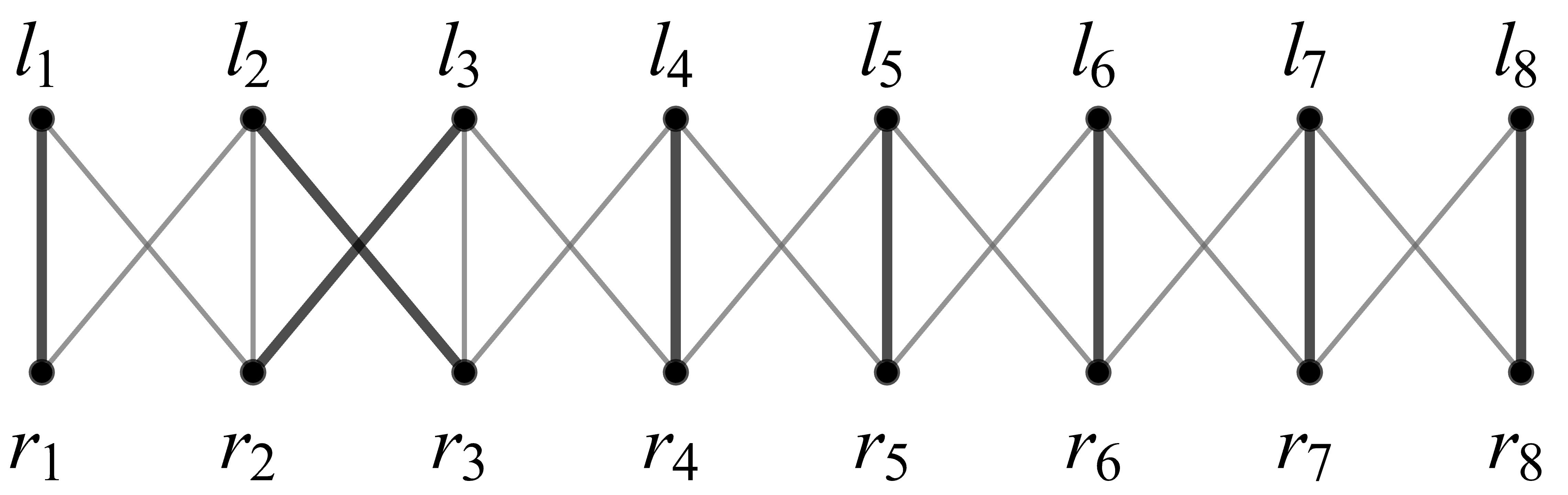}
            \label{fig:and}
        }
    }
    \hfill
    \subfigure[~Anderson with external legs \cite{Patel:2025hol}]{
        \makebox[0.48\linewidth][c]{%
            \includegraphics[
                height=0.28\textheight,
                width=0.38\linewidth,
                keepaspectratio
            ]{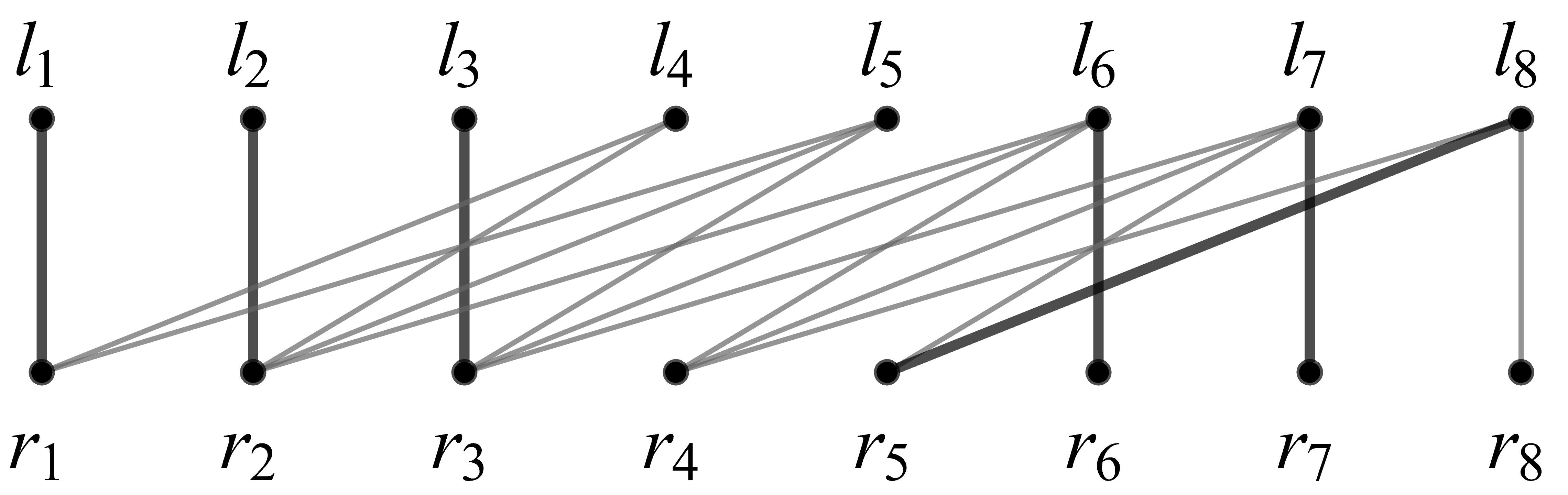}
            \label{fig:and_ext}
        }
    }
         \vspace{0.3cm}
        \subfigure[~Fractal \cite{Ibarra:2025eya}]{
        \makebox[0.96\linewidth][c]{%
            \includegraphics[
                height=0.28\textheight,
                width=0.74\linewidth,
                keepaspectratio
            ]{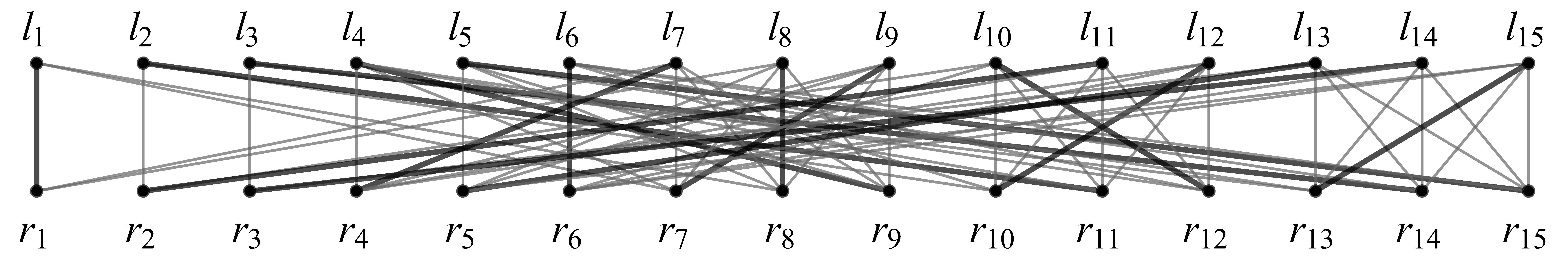}
            \label{fig:fractal}
        }
    }
    \subfigure[~Leptonic chain model \cite{Arkani-Hamed:2026wwy}]{
        \makebox[0.96\linewidth][c]{%
            \includegraphics[
                height=0.28\textheight,
                width=0.58\linewidth,
                keepaspectratio
            ]{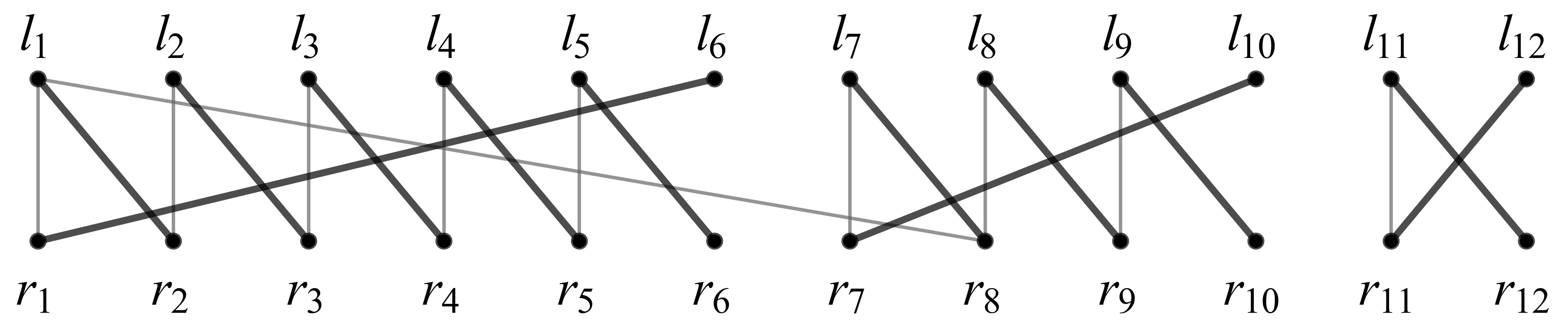}
            \label{fig:chain}
        }
    }       \vspace{0.3cm}
\caption{Maximum-matching graph representations of various latticized theory-space models. Thick edges denote elements of the maximum matching $M$. Exposed vertices are those not incident to any edge in $M$. }
\label{fig:four_panel}
\end{figure}
Perhaps the simplest and most phenomenologically well-studied example fitting into this class of scenarios is the minimal seesaw model \cite{King:1999mb,Frampton:2002qc}. It contains two right-handed neutrinos and generates non-zero masses for two of the three Standard Model neutrinos. A Dirac version of this is depicted in Fig. \ref{fig:seesaw}. The cardinality of the maximum matching in this case $\nu(G)=4$, and hence the configuration results in a massless pair of spinors following the result in Eq. (\ref{res_1}). Constructing maximum matching alternating path starting from the exposed vertices $l_2$ and $r_3$, it can be seen that the massless pair are linear combinations of $l_{1}$, $l_{2}$, $l_{3}$ and $r_{1}$, $r_{2}$, $r_{3}$ only.

The configuration corresponding to the clockwork model \cite{Giudice:2016yja} is shown in Fig. \ref{fig:clockwork} for $n=9$. This can be generalised easily to any $n$. A completely disconnected vertex is added, as the number of left- and right-chiral fields introduced in these constructions differ by one. It can be seen that $n-1$ edges can be identified as matching edges, leaving one massless mode. The latter are $l_n$ and a linear combination of all $r_i$, as can be read from $M$-alternating path configuration.

Another well-known configuration for the quantum fields in latticized theory space is allowing only the nearest-neighbour mass terms \cite{Craig:2017ppp}. This is depicted in Fig. \ref{fig:and} for $n=8$. The arrangement can give rise to localisation effects in close analogy with Anderson localisation in real lattice systems. In this case, $\nu(G)=n$ and there are no massless modes, as can be seen from Fig. \ref{fig:and}.  Interestingly, multiple massless modes can arise if additional left- and right- chiral fermions are attached to the chain on opposite ends as recently pointed out in \cite{Patel:2025hol}. Quantitatively, if $N_f$ pairs of chiral fermions are attached to a chain of $N$ fermions at the opposite ends, one finds $N_f-1$ massless modes if $2N_f-1 \leq N$. The graph-theoretic representation for $N=5$ and $N_f=3$, i.e. $n=N+N_f=8$, is displayed in Fig. \ref{fig:and_ext}. There are six edges in a maximum matching in this case, leading to two massless modes. The latter are linear combinations of $l_1,...,l_5$ and $r_4,...,r_8$ as can be read from the graph. This is in agreement with the observations made in \cite{Patel:2025hol,Mohanta:2026jnq} from analysing the full mass matrices.

We also determine the graph representation of a few recently proposed chain models. The configuration corresponding to the fractal model \cite{Ibarra:2025eya} is displayed in Fig. \ref{fig:fractal}. We find that the maximum matching is also a perfect matching in this case, leaving no massless modes arising from the geometry. The three massless modes obtained in \cite{Ibarra:2025eya}, therefore, result from the specific choice of the couplings and correlations among them.  The graph corresponding to the leptonic flavour model proposed in \cite{Arkani-Hamed:2026wwy} is shown in Fig. \ref{fig:chain}.   It can be seen as a combination of three clockwork chains with disconnected vertices matched with other vertices of the sub-chain. Therefore, this arrangement also has a perfect matching. Two of the three sub-chains are connected to give rise to a non-trivial flavour mixing.

The graph-theoretic formulation of latticized theory-space models is not merely a tool for elucidating known features; it also enables a systematic search for lattice configurations with desired properties. We illustrate this with an explicit example. Consider the model shown in Fig.~\ref{fig:and_ext}, analyzed in detail in \cite{Patel:2025hol,Mohanta:2026jnq}. It was shown that $n=8$ is the minimal number of fields required in this setup to obtain two massless modes. These modes can be identified with the first and second generations of Standard Model charged fermions, which subsequently acquire small masses through radiative corrections. Third-generation fermions naturally have masses near the electroweak scale, while the first two generations are loop-suppressed by two to four orders of magnitude. The associated vectorlike fermions and additional gauge bosons can consistently have masses of order a few TeV or higher \cite{Mohanta:2026jnq}.

\begin{figure}[t!]
\centering
\subfigure{\includegraphics[width=0.38\textwidth]{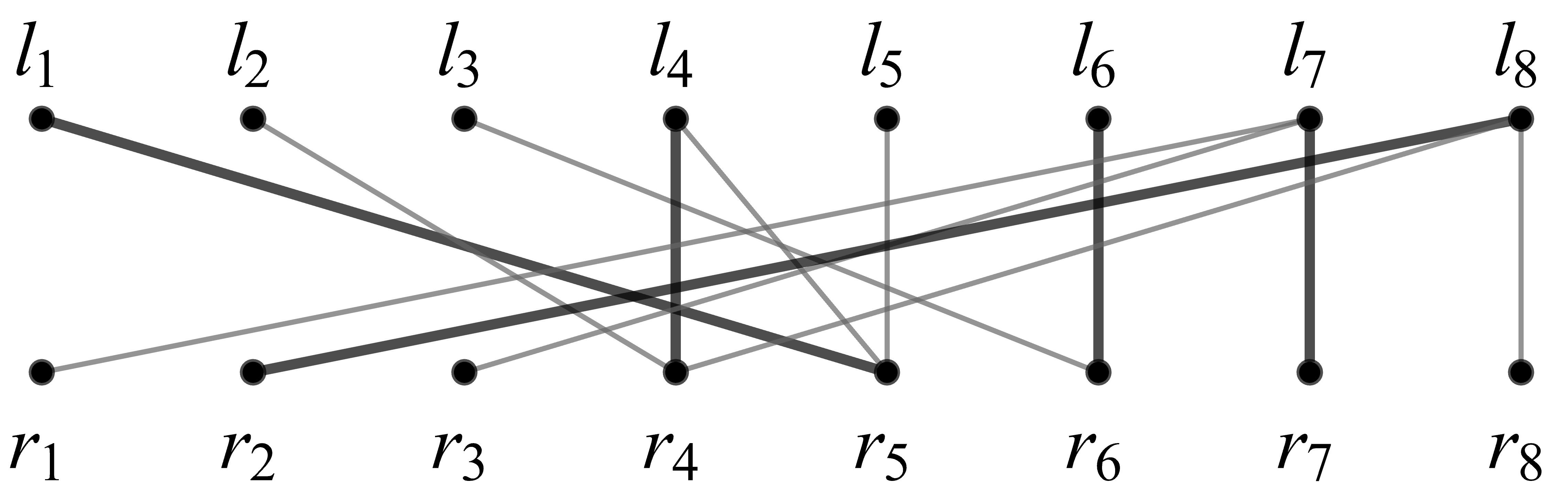}}
\caption{$n=8$ configuration with $\nu(G)=5$ leading to three massless modes at the leading order. Thick edges denote elements of the maximum matching $M$.}
\label{fig:neutrino}
\end{figure}

If the same latticized configuration is applied to neutrinos with vectorlike fermions at comparable mass scales, one generically obtains a neutrino with a mass of order the electroweak scale, in conflict with observations. It is therefore desirable that all three neutrinos be massless at leading order. The relevant question is whether, for fixed $n=8$, there exists a lattice configuration that admits three massless modes independent of the coupling strengths. This can be addressed by systematically generating candidate graphs and identifying those with $\nu(G)=3$. Using this procedure, we find an explicit example shown in Fig.~\ref{fig:neutrino}.

Upon translating the graph into a field-theoretic model, the fields $l_{1,2,3}$ can be identified with the three generations of standard neutrinos, while $r_{1,2,3}$ represent their right-chiral partners. The remaining fields constitute five Dirac fermions, which may possess bare mass terms. The required pattern of mass terms can be realised through a straightforward extension of the site-dependent global abelian symmetry introduced in \cite{Mohanta:2026jnq}. At leading order, this configuration yields three massless neutrinos, as expected.

If the vectorlike fermions additionally interact via a gauged $U(1)$ symmetry, radiative corrections generate a realistic light-neutrino mass spectrum. This mechanism is discussed explicitly in Appendix~\ref{app:nu_model}. Combined with the framework of \cite{Mohanta:2026jnq} for the charged-fermion mass hierarchy, the model provides a unified and elegant explanation of tiny neutrino masses. The length of the vectorlike fermion chain and the coupling strengths are universal across all fermion sectors. Nevertheless, the neutrino masses remain small because the vectorlike partners of the neutrinos carry comparatively small charges under the new gauge symmetry, which strongly suppresses the radiatively induced mass scale.

\section{Summary}
\label{sec:summary}
Fermion mass terms in quantum field theory can be represented by a bipartite graph, where fields correspond to vertices and nonvanishing mass terms to edges. Latticized theory space, dimensional deconstruction, and chain models are mapped to specific graph topologies. Using tools from graph theory, we establish two general results for the fermion mass spectrum: (i) the number of massless modes can be computed directly from the cardinality of a maximum matching, and (ii) the massless modes have nonvanishing overlap only with those fields that are reachable from exposed vertices via even-length alternating paths in a maximum matching. These properties are determined entirely by the graph structure and are independent of the numerical values of the parameters of the underlying models.

These observations enable a systematic construction of latticized frameworks tailored to a desired number of massless modes and specified wave-function overlaps. Several illustrative examples and applications are presented. Although the analysis focuses on fermion chains generated by Dirac mass terms, the underlying graph-theoretic framework can be generalised to other types of fields. 

\begin{acknowledgments}
The author gratefully acknowledges Gurucharan Mohanta for a careful reading of the manuscript and for helpful comments and corrections, and the anonymous referee for valuable comments and for highlighting relevant connections to existing results that helped improve this work. This research at the Physical Research Laboratory was supported by the Department of Space (DOS), Government of India.
\end{acknowledgments}

\appendix 
\onecolumngrid\
\section{Neutrino masses from radiative corrections in the latticized theory space}
\label{app:nu_model}
This appendix presents a detailed discussion of the neutrino mass generation mechanism outlined in section~\ref{sec:ex} and illustrated in Fig.~\ref{fig:neutrino}. The analysis closely follows the latticized theory-space framework developed to explain the radiative origin of charged-fermion masses in \cite{Mohanta:2026jnq}. We adopt the notation and conventions introduced there and refer the reader to \cite{Mohanta:2026jnq} for further details.

Consider three generations of lepton doublets $l_{L\alpha}$ and singlet neutrinos $\nu_{R\alpha}$ with $\alpha=1,2,3$. In close analogy with the charged-fermion construction of \cite{Mohanta:2026jnq}, we introduce five pairs of vectorlike fermions $N_{Li}$ and $N_{Ri}$ with $i=1,\ldots,5$. Only the vectorlike fermions are charged under the additional $U(1)_X$ gauge symmetry. The relevant Yukawa interactions are
\beqa \label{L:SM:Yukawa} 
-{\cal L}_Y &=& \sum_{\alpha, i} ((y_\nu)_{\alpha i}\, \overline{l}_{L \alpha} \tilde{H} N_{R i} + (y_\nu^\prime)_{i \alpha}\, \overline{N}_{L i} S^* \nu_{R \alpha}) + \sum_{i,j} \left(M^{(0)}_N\right)_{ij}\,\overline{N}_{Li} N_{Rj}\,+\, {\rm h.c.}\,, \eeqa
where $\tilde{H}$ ($S$) denotes an electroweak doublet (singlet) scalar charged under $U(1)_X$. The matrices $y_\nu$, $y_\nu^\prime$, and $M_N^{(0)}$ acquire specific textures determined by the arrangement shown in Fig. \ref{fig:neutrino}. Explicitly,
\beqa
y_\nu = \left(\ba{ccccc} 0 & y_{\nu 1} & 0 & 0 & 0 \\
y_{\nu 2} & 0 & 0 & 0 & 0 \\
0 & 0 & y_{\nu 3} & 0 & 0 \ea \right),~y^\prime_\nu = \left(\ba{ccc} 0 & 0 & 0 \\
0 & 0 & 0 \\
0 & 0 & 0 \\
y_{\nu 1}^\prime & 0 & y_{\nu 2}^\prime \\
0 & y_{\nu 3}^\prime & 0 \ea \right),~M_N^{(0)} = W_N \left(\ba{ccccc} 1 & t_N & 0 & 0 & 0\\
0 & 1 & 0 & 0 & 0\\
0 & 0 & 1 & 0 & 0\\
0 & 0 & 0 & 1 & 0\\
t_N & 0 & 0 & 0 & 1 \ea \right).
\eeqa
For simplicity, we assume universal diagonal and off-diagonal entries in $M_N^{(0)}$, although this assumption is not essential. The resulting textures differ substantially from those in the charged-fermion sector discussed in \cite{Mohanta:2026jnq}. They can be generated by imposing a global $U(1)^5$ symmetry, assigning appropriate charges to the chiral and vectorlike fermions, and introducing suitably charged flavon fields. After spontaneous breaking of the global symmetry, the flavons generate the off-diagonal entries in $M_N^{(0)}$. Further details of such constructions can be found in \cite{Patel:2025hol,Mohanta:2026jnq}.

Once $\tilde{H}$ and $S$ acquire vacuum expectation values, the full $8 \times 8$ neutrino mass matrix becomes
\be \label{calMf}
{\cal M}_\nu = \left(\ba{cc} \left(0\right)_{3 \times 3} & (\mu_\nu)_{3\times5}
\\ (\mu_\nu^\prime)_{5\times3}
& \left(M_N\right)_{5 \times 5}\ea\right)\,,\ee 
where 
\be \label{mu_f}
\mu_{\nu} = y_\nu\, \frac{v}{\sqrt{2}}\,, \indent  \mu^\prime_{\nu} = y^\prime_\nu\, \frac{v_S}{\sqrt{2}}\,, \ee
and 
\be\label{MF:1loop} 
M_N = M_N^{(0)} + \delta M^{(0)}_N\,,\ee
is the 1-loop-corrected mass matrix of vectorlike fermions. In the limit $\delta M^{(0)}_N \to 0$, the ${\cal M}_\nu$ is identical to the weighted adjacency matrix $A$ for the graph in Fig. \ref{fig:neutrino} and one finds three massless modes irrespective of the values of non-zero parameters. 

The radiative corrections induce a departure from the exact geometry and hence induce masses for these otherwise massless modes. The finite radiative corrections arise from loops involving the fermions and gauge boson of $U(1)_X$. Since only $N_{Li,Ri}$ are charged under $U(1)_X$, only $M_N^{(0)}$ gets corrected at one loop. Such corrections are computed as \cite{Patel:2025hol,Mohanta:2026jnq}
\be \label{MN_loop_corr}
\left(\delta M_N^{(0)} \right)_{ij} = \frac{g_X^2 q^2}{4 \pi^2}\, \sum_{k=1}^5\,\left(U_{NL}\right)_{ik} \left(U_{NR}\right)^*_{jk} \,m_{Nk}\,B_0[M_X^2, m_{Nk}^2]\,. \ee
Here, $g_X$ is a gauge coupling and $q$ is a charge of vectorlike fermions under $U(1)_X$. The unitary matrices $U_{NL}$ and $U_{NR}$ are such that
\be \label{MN_diag}
U_{NL}^\dagger\, M_N^{(0)}\,U_{NR} = {\rm Diag.} \left(m_{N1},...,m_{N5}\right)\,.\ee
The loop function $B_0$ is defined in \cite{Mohanta:2026jnq}. 

We find that the above can lead to a realistic neutrino mass spectrum for a reasonable choice of the values of the parameters. Following a phenomenologically viable benchmark solution from \cite{Mohanta:2026jnq}, we set $W_N=7.4$ TeV, $t_N=0.98$ and  $M_X=5$ TeV. This along with $g_X q = 10^{-4}$ and 
\beqa
\mu_\nu = \left(
\begin{array}{ccccc}
 0. & 37.58 & 0 & 0 & 0 \\
 5.86 & 0 & 0 & 0 & 0 \\
 0 & 0 & -55.09 & 0 & 0 \\
\end{array}
\right)\, {\rm GeV}\,,~\mu^\prime_\nu = \left(
\begin{array}{ccc}
 0 & 0 & 0 \\
 0 & 0 & 0 \\
 0 & 0 & 0 \\
 20.55 & 0 & -33.46 \\
 0 & -37.24 & 0 \\
\end{array}
\right)\,{\rm GeV}\,.
\eeqa
lead to 
\be \label{mnu}
m_{\nu 1} = 0\,{\rm eV}\,,~~m_{\nu 2} = 0.0865\,{\rm eV}\,,~~m_{\nu 3} = 0.05\,{\rm eV}\,,\ee
for the light neutrinos. The above reproduces the current best-fit values of the solar and atmospheric squared mass difference obtained from the global fit of neutrino oscillation data \cite{Esteban:2024eli}.

In comparison with the construction for charged-fermion mass hierarchies presented in \cite{Mohanta:2026jnq}, the number of vectorlike fermions is the same and their masses are of comparable magnitude. The principal differences between the two sectors arise from the distinct configurations of the latticized theory space and from the much weaker coupling of the neutrino vectorlike partners to the $U(1)_X$ gauge boson. The former leads to three massless modes at leading order, as opposed to two in the charged-fermion sector, while the latter ensures that neutrino masses are generated only at the sub-eV scale.

\bibliography{references}
\end{document}